# Letter to the Editor of *Sky & Telescope* Concerning Galileo's Observations of Mizar[1]

(published in *Sky & Telescope* May 2006; correction published July 2007)

Christopher M. Graney
Jefferson Community College
1000 Community College Drive
Louisville, Kentucky 40272
(502) 213-7292
christopher.graney@kctcs.edu
www.jefferson.kctcs.edu/faculty/graney

ABSTRACT:

Work published in *Sky & Telescope* in 2004 discusses Galileo's observations of the star Mizar. These observations raise questions regarding Galileo's assumptions about the universe and the conclusions he drew from his observations. Galileo would have expected Mizar to reveal annual parallax and thus provide evidence of Earth's motion, but Mizar shows no such parallax.

---

[1] Published by *Sky & Telescope* under the title "Galileo's Pride & Prejudice".

LETTER[2]

"A New View of Mizar" by Leos Ondra (July 2004, page 72) discusses how Galileo, in a search for direct evidence of Earth's motion, observed the double star Mizar in Ursa Major in an attempt to measure its parallax. However, Ondra inadvertently raises interesting questions regarding Galileo's assumptions about the universe and the conclusions that he drew from his observations.

Ondra dates Galileo's observations of Mizar to 1617. Certainly Galileo would have been very interested in parallax at the time, for Cardinal Robert Bellarmine had recently (1615) written that Catholic authorities would accept the heliocentric theory if direct evidence for Earth's motion were obtained. Ondra finds that Galileo measured the apparent diameters of Mizar's component stars to be 6 and 4 arcseconds, separated by 15 arcseconds. To determine their distances, Galileo assumed that all stars were roughly the same size as the Sun, then calculated that since Mizar A was [1/300] the apparent size of the Sun, it must be 300 a.u. distant (Mizar B would be 450 a.u. distant).

Galileo knew nothing of light from a point source diffracting through a circular aperture and couldn't know that the sizes he measured were due to wave optics and did not reflect the stars' dimensions. His size measurements would seem good, and his distance calculations would seem as good as his assumption that the stars were suns.

Galileo must have expected the components of Mizar to swing around each other dramatically as he observed them over a period of weeks and months. Based on his calculations he would have expected A and B to have parallax angles of ±11.5 and ±7.6 arcminutes, respectively. Their relative motion would dwarf their separation. But in fact Mizar A and B do not budge. Since no parallax is seen, Galileo logically had to conclude either that the Earth was stationary or that his assumption regarding stars being suns at differing distances from Earth was wrong.

---

[2] *Sky & Telescope* added an illustration to the letter -- a portrait of Galileo by D. Robusti. The caption for this illustration (also added by *S&T*) reads "Galileo, along with his student Benedetto Castelli, was the first to discover and observe binary stars. But what did those observations reveal about Galileo's scientific integrity? This portrait was made by Domenico Robusti around 1605-07."

Yet Galileo asserts both these things in his *Dialogue Concerning the Two Chief World Systems* (1632). In the *Dialogue,* Galileo argues[3] that 1st-magnitude and 6th-magnitude stars have apparent sizes of 5 and 5/6 arcsecond, respectively, that stars are the same size as the Sun, and that since 5/6 arcsecond is 1/2160 the size of the Sun, 6th-magnitude stars are 2160 a.u. distant — arguments in line with his work on Mizar. He states, "If some tiny star were found by the telescope quite close to some of the larger ones, and if that one were therefore very remote, it might happen that some sensible alterations would take place among them," and he goes on to suggest that the "sensible alterations," or parallax, would provide proof of Earth's motion.

The *Dialogue* conflicts with Galileo's earlier work on Mizar, raising interesting points.[4] Had Galileo published his Mizar observations, they would have influenced the ongoing debate regarding Earth's motion, likely prolonging the time before the geocentric theory was finally overturned. It would seem that at the time the *Dialogue* appeared, Galileo was sitting on results that strongly challenged the Copernican theory he was championing!

---

[3] In the published version in *S&T* this reads "Yet Ondra reminds us that in *Dialogue Concerning the Two Chief World Systems* (1632), Galileo argues…." This change was made after the author reviewed the final proof. The author did not ask S&T to publish a correction regarding this change.

[4] In the published version in *S&T* this reads "The Dialogue conflicts with Galileo's earlier work on Mizar, raising some very interesting ethical points." This change was made after the author reviewed the final proof. The author asked *S&T* to publish a correction regarding this change, which *S&T* agreed to do. The correction was published in the July 2007 issue of *S&T*.